\documentclass[aps,twocolumn,prl]{revtex4}
\usepackage{epsfig}
\begin{document}

\title{On Graphene Hydrate}
\author{
Wei L. Wang and Efthimios Kaxiras
}
\affiliation{
Department of Physics and School of Engineering and Applied Sciences,
Harvard University, Cambridge, MA 02138
}
\date{\today}

\begin{abstract}
Using first-principles calculations, we show that the formation of carbohydrate directly from carbon and water is energetically favored when graphene membrane is subjected to aqueous environment with difference in chemical potential across the two sides. The resultant carbohydrate is two-dimensional, where the hydrogen atoms are exclusively attached on one side of graphene while the hydroxyl groups on the other side form a herringbone reconstruction that optimizes hydrogen bonding. We show that graphene undergos semi-metal-insulator transition upon hydration which is readily detectable from the significant shift in the vibration spectrum. The hydrate form of graphene suggests new applications for graphene in electronics, either deposited on a substrate or in solution.
\end{abstract}

\pacs{
68.65.Pq, 
68.43.Bc  
82.65.+r  
73.61.Wp  
}

\maketitle

Carbohydrate, C$_m$(H$_2$O)$_n$, is of great importance since it plays a crucial role in various life functions, most notably as the medium for
the storage and transport of energy in living systems. In the
structure of carbohydrate,  the protons and the hydroxy groups separately bond to each carbon atom.
It is an intriguing possibility to create a macroscopically large, planar carbohydrate structure based on
graphene, a newly discovered carbon allotrope~\cite{Novoselov2004} that has
excited much interest for novel two-dimensional (2D)
material structures and electronic device applications~\cite{Geim2009}.
Recent theory~\cite{Sofo2007} and experiment~\cite{Elias2009} showed that it is possible to hydrogenate graphene at room-temperature to produce graphane, a process which is reversible.
In organic chemistry, on the other hand, it is readily possible
to insert water molecules into alkene species through
hydration reactions, for example, converting ethylene to ethanol.
Both these processes involve opening the double C-C bond so that protons or
the hydroxy group can be attached to C atoms.
Hydration of graphene, however, is elusive because its planar geometry puts severe
constraints on the final configuration: if the proton and hydroxy groups are on the same
side of the C-atom plane, the 2D periodicity of the lattice places them so
close that they recombine to form water molecules.
Therefore, hydration of graphene is not feasible unless the protons and hydroxy groups
can be placed on opposite sides of the C-atom plane.

These considerations have motivated our study of what would emerge if graphene could be
subjected to two very different chemical environments, one on each side.
The difference in chemical environment could be the ion concentration,
pH value or ionic potential. The result of this process may open new possibilities for
device applications of graphene in solution, for instance, single DNA sequencing
through ionic transport and novel battery technology based on graphene membranes.
In this letter, we examine the properties of a 2D crystalline hydrate
structure obtained from graphene, where the hydroxy groups spontaneously
lock into a herringbone structure on one side of the basal C-atom plane
while protons are chemisorbed on the other.
We show that this structure, which we refer to as ``graphanol'' or ``poly-ethanol'',
is stable at room temperature. We notice that a similar structure in analogy to siloxane was recently studied~\cite{Nakamura2009} but a small unit cell was used which prevents the system
from reconstructing to the most favorable configuration.
We show that upon hydration, graphene which is a semi-metal in pure form,
transforms into a insulator with a substantial direct gap.
Due to charge transfer and alignment, a macroscopic net dipole moment
arises perpendicular to the C-atom plane.
Hydration has dramatic impact on the vibrational spectra, with new peaks arising
and the bond stretching mode of graphene shifting to lower energy.

For our study of graphanol, we use first-principles calculations
based on density functional theory. The geometry and formation energies were calculated
using projector-augmented plane waves as implemented in VASP~\cite{Kresse1996}
with an energy cutoff of 400 eV. Structure optimization stops when the magnitude of the force
on each atom is less than 0.04 eV/{\AA}, and stress on the cell is less than 0.01 GPa.
The optimized geometry was checked with the SIESTA code~\cite{Soler2002}
which employs local atomic orbitals, with an energy cutoff of 70 Ry.
The lattice constants obtained from the two approaches differ less than $1\%$.
The local orbital method was also used for the calculation of the projected density of states (DOS),
density of charge difference and molecular dynamics (MD) simulations.
In the latter, a $4 \times 4$ supercell of graphene was used, with a time-step of 1 fs
and constant temperature was maintained by a Nos\'{e} thermostat.
In all calculations, the PBE exchange-correlation functional and a
$4 \times 4 \times 2$ k-point mesh were used unless otherwise specified.

\begin{figure}
\includegraphics[width=0.45\textwidth]{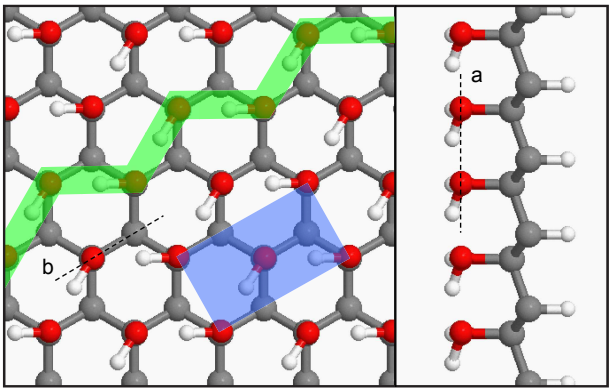}
\caption{Top view (left) and side view (right) of the herringbone structure of graphene hydrate with
protons (white) and hydroxy groups (oxygen in red), residing exclusively on one side of the membrane,
both groups chemisorbed on carbon atoms (gray).
One of the zigzag herringbone lines (green stripe) and a unit cell (blue rectangle) are highlighted
in the top view. The dashed lines marked as ``a'' and ``b'' denote the slicing planes in Fig. 3a and 3b .}
\label{fgr:1}
\end{figure}

\begin{figure}
\includegraphics[width=0.45\textwidth]{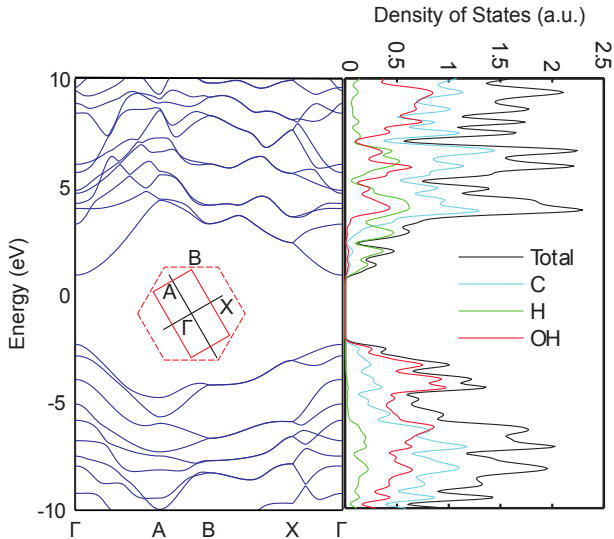}
\caption{Band structure (left) and density of states projected to atomic orbitals (right).
The first Brillouin zone of graphanol (solid red) and graphene (dashed red) are shown as inset,
with high-symmetry points labeled.}
\label{fgr:2}
\end{figure}

Upon hydration, with the protons and hydroxy groups chemisorbed on each side
of graphene (Fig. 1), the planar graphene structure is converted to a
puckered diamond-like sp$^3$ structure with the C-C bond length increasing
from 1.42 {\AA} to 1.56--1.57 {\AA}. The in-plane lattice constant increases by $5\%$
compared to that of graphene.
The C-C bond elongation is expected because of the lost of the $\pi$ bonds.
The C-H and C-O bond lengths are 1.11 {\AA} and 1.43 {\AA}, respectively, typical values for
hydrocarbons and alcohols.
In principle, the hydroxy group can rotate freely about the C-O bond;
our MD simulation at 300K however shows that these groups spontaneously form a herringbone structure in which the proton in one of the hydroxy groups points to one of its neighboring
oxygen atoms. This structure changes the periodicity of the 2D crystal to a rectangular lattice,
as shown in Fig. 1.
Compared to the structures with periodicity restricted to that of the graphene unit cell,
the formation of the herringbone structure lowers the energy by about 126 meV per hydroxy group,
which is in the range of typical hydrogen bond energy. The overall binding energy indicates that the system is more stable than its isomer ketene (C$_2$H$_2$O) and diketene (C$_4$H$_4$O$_2$) by 0.97 eV and 0.43 eV per C$_2$H$_2$O unit respectively.
Our MD simulation, run for 3 ps, shows that the system remains stable at 800K.

Similar to hydrogenation of graphene, the hydration of graphene leads to a semi-metal -- insulator
transition of pristine graphene, as is shown in Fig. 2. Our calculation shows that
graphanol with the herringbone hydroxy surface has a direct gap of 3.22 eV,
which is slightly less than that of graphane. The plot of the density of states projected to local atomic orbitals (Fig. 2, right panel) reveals that the top of the valence band predominantly resides on the hydroxy group and the carbon atoms while the bottom of the conduction band
is related to the single hydrogen and the carbon atoms.
This indicates that excitation of the system would be accompanied by a transfer of
electronic charge across the
C-atom plane, from the hydroxy to the hydrogen terminated side.

\begin{figure}
\includegraphics[width=0.45\textwidth]{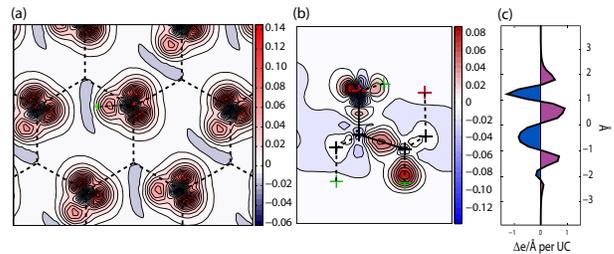}
\caption{(a) and (b) Charge difference distribution (units of e/{\AA}$^3$) of the system shown on the planes indicated in Fig. 1 as lines ``a'' and ``b''.
(c) Integrated charge difference along the direction perpendicular to the C-atom plane. The colored crosses mark the projected position of the atoms (green for H, red for O, and gray for C) while the in-plane (out-of-plane) bonds are marked with solid (dashed) lines.}
\label{fgr:3}
\end{figure}

To understand the bonding nature of the system, we show in Fig. 3 the charge density difference
between the graphanol structure and the superposition of charge of the individual free atoms
placed at the same positions as in graphanol.
As is clear from these plots, the herringbone reconstruction is driven by the hydrogen bond
interaction between the depleted proton $s$ orbital and one of the lone pair orbitals of the oxygen.
The attraction between these two requires that the OH bond point in the direction of
a neighboring oxygen whose OH bond in turn rotates away from that direction by $60^0$
so that one of its lone pairs engages in strong interaction with the proton, along the zigzag line.
This pattern matches the six-fold symmetry of graphene and results in a
herringbone structure across the 2D crystal.
Both C atoms in the unit cell of the herringbone structure (Fig. 3b) have clear $sp^3$ hybridized
orbitals which form covalent bonds to the oxygen atoms on one side and to the
hydrogens on the other.
The charge difference is integrated in the plane of the membrane and plotted in Fig. 3c along the
direction perpendicular to the plane.
This leads to a net dipole moment of 0.2 e{\AA} per unit cell, pointing toward the single protons.
Compared to the dipole of a water molecule projected in one OH bond direction,
this value is reduced by a factor of 2, due to the separation of the positive and negative charges
that form the dipole, by the C-atom plane, which partially screens their interaction.
Remarkably, a macroscopic dipole moment should arise since all OH$^-$ -- H$^+$ units
in the hydrate are aligned in the perpendicular direction as long as the C-atom plane
remains flat.

\begin{figure}
\includegraphics[width=0.45\textwidth]{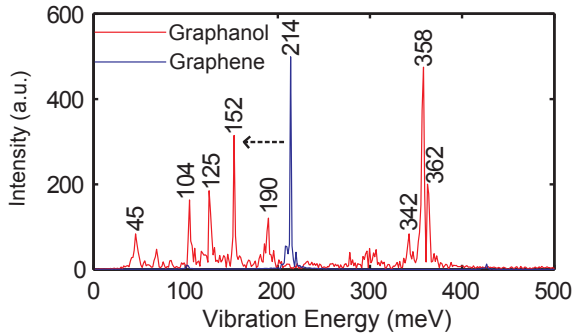}
\caption{Calculated vibrational spectra (at $\Gamma$, the center of the Brillouin Zone)
of pristine graphene (blue) and that of graphene upon hydration (red).}
\label{fgr:4}
\end{figure}

Because of the thinness of the graphene plane,
the vibrational properties of the single atomic layer change dramatically upon hydration.
We calculated the vibrational spectra of both graphene and graphanol using first-principles
MD simulation runs for 3 ps.
The vibrational modes are extracted by Fourier transforming the velocity autocorrelation function
along the perpendicular direction. The C-C bond stretching mode is found red shifted
from 214 meV for graphene to 152 meV for graphanol due to the added mass of the adsorbates.
On the other hand, a new prominent higher energy peak arises at 358 meV which corresponds to the
C-H bond stretching mode. Other peaks at 342, 190, and 104 meV, respectively,
correspond to the O-H bond stretching mode, the O-H bond scissor mode, and
the C-H bond scissor mode.
For complete details and movies illustrating the atomic displacements
that correspond to these modes, see supporting information.

The herringbone graphanol structure we considered here is metastable.
Its formation from graphene and H$_2$O is endothermal by 0.78 eV per C$_2$H$_2$O unit.
This holds true also for graphane compared to graphene and H$_2$, which has been
experimentally observed to be stable at room temperature.
Graphanol is more stable than its known isomers as we already mentioned,
and remains stable at elevated temperatures as indicated by our MD simulations.
The experimental synthesis of graphane involved low energy plasma which provides
the kinetic energy to overcome the barrier for chemisorption.
For graphanol, the adsorbates on each side of the C-atom plane, namely the proton and
hydroxy groups, are opposite in nature. We suggest that graphanol can be synthesized in solution
by controlling the pH value so that protons dominate on one side of the graphene membrane and the hydroxide anions dominate on the other. The formation of graphanol from graphene, proton and hydroxide anions is strongly exothermic, releasing 2.56 eV per C$_2$H$_2$O unit.
One way to control this reaction in experiment is to apply a potential difference across the membrane in water, a setup that would establish the concentration difference and also a strong electric field across the membrane that helps overcome the barrier of chemisorption.
Furthermore, in such a scheme, the reaction can likely be reversed by switching the direction
in which the potential difference is applied. Indeed, recent experiments~\cite{Echtermeyer2008}
observed such reversible conducting-to-insulating behavior in a gated graphene device and the
origin of the transition was attributed to water.

\begin{figure}
\includegraphics[width=0.45\textwidth]{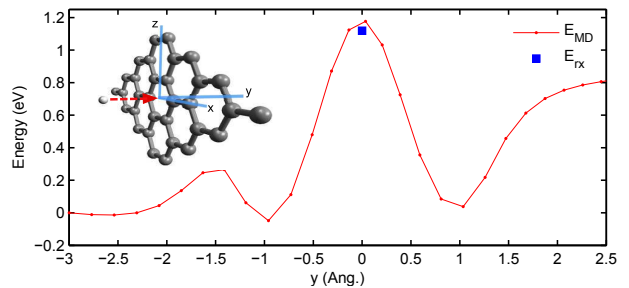}
\caption{Estimation of the energy barrier for penetration of graphene by a proton.
The proton (white sphere in the inset) moves through the center of a hexagonal ring
in graphene with a minimum starting kinetic energy.
The energy profile (red line) is the potential energy of the system
(total energy minus the kinetic energy of the ions) at various proton-graphene distances
in the MD simulation.
The blue square marks the energy when the structure is relaxed with the proton
fixed at the center of the hexagonal ring of graphene, giving a barrier of 1.17 eV.
}
\label{fgr:5}
\end{figure}

An obvious question in the proposed graphene hydration scheme is whether ions of one type can
penetrate the graphene membrane and end up on the wrong side.  The hydorxyl ions are too large
to achieve this, given the size of holes on the grahene layer, since the hexagonal carbon network of pristine graphene is surprisingly dense and impenetrable to small gas molecules such as helium~\cite{Bunch2008}.
The only relevant issue then is the
possible penetration of the membrane by the protons.  In order to address this we have
performed MD simulations to establish the energy barrier for proton penetration.
Fig. 5 shows the energy profile when a proton moves through graphene.
The proton starts with a kinetic energy that is barely enough to penetrate
graphene at the center of the hexagonal ring. The highest point of the profile corresponds to an upper bound of the required system energy for the penetration.
We estimate a lower bound of the barrier by relaxing the structure with the proton held fixed
at the center of the hexagonal ring. The barrier found is 1.17 eV.
For hydrogen and larger species such as the hydroxy group the penetration barrier is much higher. These barriers prevent protons or hydroxide anions from diffusing to the other side of graphene and recombine to form water instead of forming graphanol. It is worth pointing out that because of the strong interaction of the proton and hydroxy group across the atomically thin membrane,
the formation of graphanol is favored (by 2.25 eV per C$_2$H$_2$O unit)
over formation of hydrogen and oxygen gas,
in contrast to the situation in a bulk metal, for which hydrolysis of water could occur.

The graphene hydration process described above represents a unique case where the reaction
conditions are directly manipulated on a sub-nanometer length scale.
Considering graphene as a giant flat molecule, a view originally suggested by Pauling~\cite{Geim2009},
this reaction would consist of assembling a different type of a single macromolecule where
the two ion species on each side of the C-atom plane interact strongly. Graphene is a mostly inert, strong and atomically thin material of semi-metallic character. Its range of applications could be broadly extended by introducing its insulating counterpart, the graphanol structure discussed here.
For instance, such a thin membrane could prove very useful in studies of electronic DNA sequencing
in ionic solution, where an insulating membrane could be essential in confining the
electronic tunneling current to the nucleic bases for an easily detectable signal.
Other possible applications may include electronic devices in which both the
conducting and insulating parts are composed of single planes of C atoms,
patterned from the same piece of graphene
to form areas of pure graphene and graphanol, as required by the circuit design.

\end{document}